\documentclass[12pt,twoside]{article}

\date{}

\usepackage{natbib}
\usepackage {graphicx} 

\begin{document}
\centerline{\bf Adv. Studies Theor. Phys, Vol. x, 200x, no. xx,
xxx - xxx}

\centerline{}

\centerline{}

\centerline {\Large{\bf The relativistic equation of motion in  }}

\centerline{}

\centerline{\Large{\bf turbulent jets}}

\centerline{}

\centerline{\bf {L. Zaninetti}}

\centerline{}

\centerline{Dipartimento  di Fisica Generale}

\centerline{Universit\`a degli Studi di Torino}

\centerline{via P.Giuria 1,  10125 Torino,Italy}

\begin{abstract}
The turbulent jets are usually described by classical velocities.
The relativistic case can be treated  starting from the
conservation of the relativistic momentum. The two key assumptions
which
 allow to obtain a simple expression for the relativistic
trajectory and relativistic velocity are null pressure and
constant density.
\end{abstract}

{\bf PACS:} 47.75.+f ; 47.27.-i; 47.27.wg \\
{\bf Keywords:} Relativistic fluid dynamics ; Turbulent flows
        Turbulent jets

\section{Introduction}
The theory of turbulent jets emerging from a circular hole can be
found in different books which adopt different theories
\cite{foot,landau,goldstein,Pope2000} or in specialized articles
\cite{List1982,Launder1983} : in  all the models the pressure ,
$p$, is equal to zero. The centerline velocity is always taken to
be classical. This paper briefly reviews in
Section~\ref{classical} the simplest model  of turbulent jets and
then investigates in Section~\ref{relativistic} the case of
relativistic centerline velocity .

\section{Classical equations}
\label{classical}

The starting point is the conservation of the momentum's flux in a
"turbulent  jet"  as outlined in  \cite {landau} (pag.~147). The
section ,$A$, is  :
\begin {equation}
A=\pi~r^2
\quad
\end{equation}
where $r$ is the radius of the jet.
Once $\alpha$ the opening angle ,
$x_0$ the initial position on the $x$--axis and
$v_0$ the initial velocity are introduced, the
 section  $A$ at  position $x$  is
\begin {equation}
A(x)=\pi \bigl ({\it r_0}+ \left( x-{\it x_0} \right) \tan \left(
(1/2)\,\alpha
 \right) \bigr )^2
\quad  .
\end{equation}
The conservation  of the total momentum flux states that
\begin{equation}
\rho  v_0^2 A_0 =
\rho  v(x)^2 A(x)
\quad  ,
\label{conservazione}
\end {equation}
where $v(x)$ is the velocity at  position $x$
and $A_0$ the initial section.

Due to the turbulent transfer, the density $\rho$
is the same on both the two sides of
equation~(\ref{conservazione}).
The trajectory of the jet  as a function of the time
is easily  deduced
from equation~(\ref{conservazione})
\begin {eqnarray}
x= -{\frac {-{\it x_0}\,\tan \left( (1/2)\,\alpha \right) +{\it r_0}-
\sqrt {{\it r_0}\, \left( {\it r_0}+2\,\tan \left( (1/2)\,\alpha
 \right) {\it v_0}\,t \right) }}{\tan \left( (1/2)\,\alpha \right)
}} \quad  .
\label{traiettoriac}
\end{eqnarray}

The  velocity turns out to be
\begin{equation}
{\it v(t)}={\frac {{\it v_0}\,{\it r_0}}{\sqrt {{\it r_0}\, \left(
{\it r_0} +2\,\tan \left( (1/2)\,\alpha \right) {\it v_0}\,t
\right) }}} \quad  . \label{velocitac}
\end {equation}

\section{Relativistic equations}

\label{relativistic}

A relativistic flow on flat space time  is described by
the energy-momentum tensor  ,$ T^{\mu\nu} $,
\begin{equation}
T^{\mu\nu} = w u^{\mu} u^{\nu} - p g^{\mu\nu}
\quad  ,
\end{equation}
where $u^{\mu}$ is the 4-velocity ,
and the Greek  index  varies  from 0 to 3 ,
$w$ is the enthalpy for unit volume ,
$p$ is the pressure and
$g^{\mu\nu}$ the inverse metric of the manifold
\cite{landau,Hidalgo2005,Gourgoulhon2006}.
The  momentum conservation
in the presence of velocity , $v$, along one direction
states that
\begin{equation}
(w (\frac{v}{c})^2 \frac { 1}{ 1 -\frac {v^2}{c^2} } +p) A = cost
\quad ,
\label{enthalpy}
\end{equation}
where $A(x)$ is the considered area in the direction perpendicular
to the motion. The enthalpy for unit volume is
\begin{equation}
w= c^2 \rho  + p
\quad ,
\end{equation}
where $\rho$ is the   density ,
and $c$ the light velocity.
The reader  may be puzzled by the
$\gamma^2$ factor in equation~(\ref{enthalpy}),
where $\gamma^2= \frac { 1}{ 1 -\frac {v^2}{c^2} }$.
 However it should be remembered  that
$w$ is not an enthalpy, but an enthalpy per unit volume:
the extra $\gamma$
factor arises from  "length contraction" in the direction of motion
\cite{Gourgoulhon2006}.
According to the current models on classical turbulent jets
we  insert $p=0$  and
the momentum conservation law
is
\begin{equation}
(\rho  v^2 \frac { 1}{ 1 -\frac {v^2}{c^2} }) A = cost
\quad .
\end{equation}

Note the similarity between the previous formula and
condition (135.2) in \cite{landau} concerning
the shock waves : when $A$=1 they are equals.
We assume  that the cross section of the relativistic jet
grows as
\begin{equation}
A(x) = \pi \times ( r_0 + \tan ( \frac{\alpha}{2}))^2
\quad ,
\end{equation}
with $\alpha$ ,
the opening angle  of the jet, constant.
In two sections of the jet we have :
\begin{eqnarray}
\rho\,{{\it v_0}}^{2}\pi \,{{\it r_0}}^{2}{\frac {1}{ 1-{\frac {{{
\it v_0}}^{2}}{{c}^{2}}}}}&=  \nonumber  \\
\rho\,{v}^{2} \left( (\pi \,{{\it r_0}}^{2}+
\pi \,{\it r_0}\,x\alpha+O \left( {\alpha}^{2} \right) ) \right) {
\frac {1}{ 1-{\frac {{v}^{2}}{{c}^{2}}}}} &~
\quad ,
\label{conservazionerel}
\end {eqnarray}
where $v$ is the velocity at  position $x$, $v_0$ the velocity
at $x$=0 and  $c$ the light velocity.

As  due to the turbulent transfer, the density $\rho$
is the same on both  sides of
equation~(\ref{conservazionerel}) and the  following
second degree equation in $\beta=\frac{v}{c}$ is obtained:
\begin{eqnarray}
{{\it \beta}}^{2}{\it r_0}+{{\it \beta}}^{2}x\alpha-{{\it \beta}}^{2}{{
\it \beta_0}}^{2}x\alpha-{{\it \beta_0}}^{2}{\it r_0}=0
\end{eqnarray}
where $\beta_0= \frac{v_0}{c}$.
The positive solution is :
\begin{equation}
\beta=
{\frac {\sqrt {{\it r_0}}{\it \beta_0}}{\sqrt {{\it r_0}+x\alpha-{{\it
\beta_0}}^{2}x\alpha}}}
\quad .
\label{beta}
\end{equation}
From equation~(\ref{beta}) it is possible to deduce the distance
$x_{f}$ after
which the velocity is a fraction $f$  of the  initial value,
$v=f v_0$ :
\begin{equation}
 x_{f} =
\frac {
{\it r_0}\, \left( {f}^{2}-1 \right)
}
{
\alpha\,{f}^{2} \left( -1+{{\it \beta_0}}^{2} \right)
}
\quad .
\label{betaf}
\end {equation}

The trajectory of the relativistic jet  as a function of the time  can  be
deduced from equation~(~\ref{beta}) and is
\begin {equation}
\int _{0}^{x}
\frac{ 1}{{\frac {\sqrt {{\it r_0}}{\it \beta_0}}{\sqrt {{\it r_0}+x\alpha-{{\it
\beta_0}}^{2}x\alpha}}}
} dx =ct
\quad  .
\end{equation}
On integrating    the equation of the trajectory
is obtained
\begin{equation}
\frac{2}{3}\,{\frac {{{\it r_0}}^{3/2}- \left( {\it r_0}+x\alpha-{{\it \beta_0}}^{
2}x\alpha \right) ^{3/2}}{\alpha\, \left( -1+{{\it \beta_0}}^{2}
 \right) \sqrt {{\it r_0}}{\it \beta_0}}}-ct=0
\quad .
\label{firstreltraj}
\end{equation}
After some manipulation equation~(\ref{firstreltraj})
becomes  a cubic polynomial equation
\begin{eqnarray}
 \left( -3\,{\alpha}^{3}{{\it \beta_0}}^{2}+3\,{\alpha}^{3}{{\it \beta_0}}
^{4}-{{\it \beta_0}}^{6}{\alpha}^{3}+{\alpha}^{3} \right) {x}^{3}+
  \nonumber\\
 \left( 3\,{\it r_0}\,{\alpha}^{2}-6\,{\it r_0}\,{\alpha}^{2}{{\it \beta_0
}}^{2}+3\,{\it r_0}\,{{\it \beta_0}}^{4}{\alpha}^{2} \right) {x}^{2}+
 \nonumber\\
 \left( -3\,{{\it r_0}}^{2}{{\it \beta_0}}^{2}\alpha+3\,{{\it r_0}}^{2}
\alpha \right) x   -
 \nonumber\\
9/4\,{c}^{2}{t}^{2}{\alpha}^{2}{\it r_0}\,{{\it \beta_0}
}^{6}-3\,{{\it r_0}}^{2}ct\alpha\,{\it \beta_0}+3\,{{\it r_0}}^{2}ct\alpha
\,{{\it \beta_0}}^{3}-      \nonumber\\
9/4\,{c}^{2}{t}^{2}{\alpha}^{2}{\it r_0}\,{{\it
\beta_0}}^{2}+9/2\,{c}^{2}{t}^{2}{\alpha}^{2}{\it r_0}\,{{\it \beta_0}}^{4}
=0
\quad .
\label{reltraj}
\end{eqnarray}

We briefly review that in order to  solve the cubic polynomial equation
\begin{equation}
a_{{0}}{x}^{3}+a_{{1}}{x}^{2}+a_{{2}}x+a_{{3}}
=0
\quad ,
\label{cubicequation}
\end{equation}
 for $x$,
the first step is to apply the  transformation
\begin{equation}
x=y- \frac{1}{3}\,{\frac {a_{{1}}}{a_{{0}}}}
\quad .
\label{transxy}
\end{equation}
This reduces the equation to
\begin{equation}
y^3 + py + q = 0
\quad ,
\label{cubicequation_reduced}
\end{equation}
where
\begin{eqnarray}
p & = & \frac {1}{3}\,{\frac {3\,a_{{2}}a_{{0}}-{a_{{1}}}^{2}}{{a_{{0}}}^{2}}}     \\
q & = & \frac {1}{27}\,{\frac
{27\,a_{{3}}{a_{{0}}}^{2}+2\,{a_{{1}}}^{3}-9\,a_{{2}}a_{{
1}}a_{{0}}}{{a_{{0}}}^{3}}} . \\
\end{eqnarray}
The next step is to
compute the first derivative  of  the  left hand side
of equation~(\ref{cubicequation_reduced}) calling it $f(y)^{\prime}$
\begin{equation}
f(y)^{\prime}= 3y^2 + p
\quad .
\label{firstderivate}
\end{equation}
In our case  $p = 0 $  and  in the range of existence $- \infty <
y < \infty $   the first derivative is always positive and
equation~(\ref{cubicequation_reduced}) has  only one root which is
real , more precisely ,
\begin{eqnarray}
y=\frac{1}{6}\,\sqrt [3]{-108\,{\it q}+12\,\sqrt {12\,{{\it p}}^{3}+81\,{{\it
q}}^{2}}}   &  ~\nonumber \\
-2\,{\frac {{\it p}}{\sqrt [3]{-108\,{\it q}+12\,\sqrt {12
\,{{\it p}}^{3}+81\,{{\it q}}^{2}}}}}
\quad ,                      &~
\label{pqsolution}
\end{eqnarray}
or equation~(\ref{cubicequation}) has the solution
in terms of $a_0$,$a_1$,$a_2$ and $a_3$
\begin{eqnarray}
x=
\frac{1}{6}\,\sqrt [3]{
A
+12\,\sqrt {
B
}} & ~\nonumber \\
- \frac{2}{3}\, \left( 3\,a_{{2}}a_{
{0}}-{a_{{1}}}^{2} \right)
\frac {1} {a_0^2} {\frac {1}{\sqrt [3]{
A
+12\,\sqrt {B} }}} &
\label{asolution}
\end{eqnarray}
where
\begin{eqnarray}
A=-36\,{\frac {3\,a_{{2}}a_{{0}}-{a_{{1}}}^{2}}{{a_{{0}}}
^{2}}}   \nonumber    \\
B=\frac{4}{9}\,{\frac { \left( 3\,a_{{2}}a_{{0}}-{a_{{1}}}^{2}
 \right) ^{3}}{{a_{{0}}}^{6}}}+9\,{\frac { \left( 3\,a_{{2}}a_{{0}}-{a
_{{1}}}^{2} \right) ^{2}}{{a_{{0}}}^{4}}}  \nonumber
\quad .
\end{eqnarray}

When the equation~(\ref{reltraj}) is considered we have  $p=0$ and
therefore we have one real  root which  is  :
\begin {eqnarray}
x(t)= \nonumber &~\\
  \frac{1}{2}{\frac {2{\it r_0}-\sqrt [3]{2}\sqrt [3]{{\it r_0}} \left( 2{
\it r_0}+3ct\alpha{\it \beta_0}-3ct\alpha{{\it \beta_0}}^{3}
 \right) ^{2/3}}{\alpha \left( {{\it \beta_0}}^{2}-1 \right) }}
\label{traiettoriarel}
& .
\end{eqnarray}
This is the law of motion of the
relativistic turbulent jets and the velocity
as function
of the time is
\begin{equation}
v(t) =
{\frac {c{\it \beta_0}\sqrt [3]{{\it r_0}}\sqrt [3]{2}}{\sqrt [3]{2{
\it r_0}+3ct\alpha{\it \beta_0}-3ct\alpha{{\it \beta_0}}^{3}}}}
\quad .
\label{velocitar}
\end{equation}

\begin{figure}
  \begin{center}
\includegraphics[width=8cm]{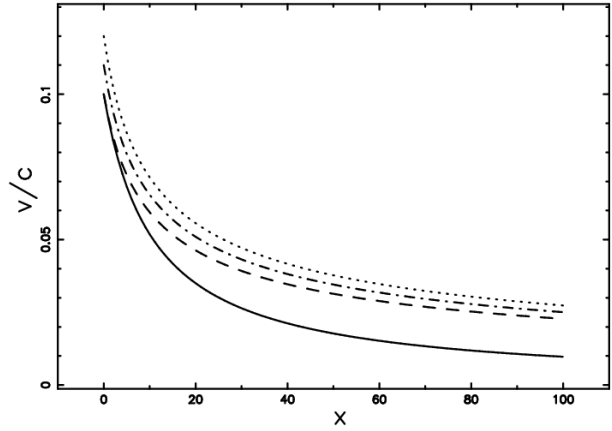}
  \end {center}
\caption {
 Velocity as function of the distance from the nozzle
 when  $r_0=1$   , $x_0=0 $, $\alpha =0.185$
 and
 $v_0/c =0.1        ~(classical ~case,~equation~(\ref{velocitac})) $
       (full line        )
 ,
 $v_0/c=\beta_0 =0.1~(relativistic ~case,~equation~(\ref{velocitar})) $
   (dashed           )
,
 $v_0/c=\beta_0 =0.11~(relativistic ~case,~equation~(\ref{velocitar})) $
   (dot-dash-dot-dash          ) and
 $v_0/c=\beta_0 =0.12~(relativistic ~case,~equation~(\ref{velocitar})) $
( dotted ).
          }%
    \label{velocity}
    \end{figure}

\begin{figure}
  \begin{center}
\includegraphics[width=8cm]{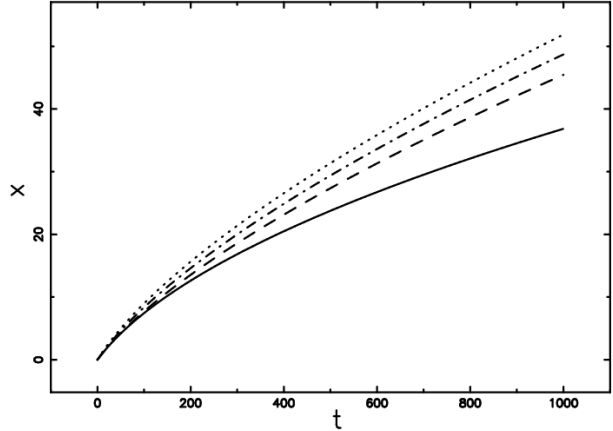}
  \end {center}
\caption {
 Distance from the nozzle as function of the time
 when  $r_0=1$   , $x_0=0 $, $\alpha =0.185$ , $c=1$
 and
 $v_0/c =0.1        ~(classical ~case,~equation~(\ref{traiettoriac})) $
 (full line        ),
 $v_0/c=\beta_0 =0.1~(relativistic ~case,~equation~(\ref{traiettoriarel})) $
 (dashed           ),
 $v_0/c=\beta_0 =0.11~(relativistic ~case,~equation~(\ref{traiettoriarel})) $
 (dot-dash-dot-dash) and
 $v_0/c=\beta_0 =0.12~(relativistic ~case,~equation~\ref{traiettoriarel})) $
    (dotted).
          }%
    \label{trajectory}
    \end{figure}

Figure~\ref{velocity} reports the classical and relativistic
behavior of the velocity as a function of the distance from
the nozzle and Figure~\ref{trajectory} the distance traveled
by the jet as a function of the time.

\section{Conclusions}

The complicate behavior  of the energy tensor that describes the
relativistic fluids takes a simple expression in the case of null
pressure. This allows to deduce the law of motion of the
relativistic turbulent jet as well the velocity behavior as
function of the distance from the nozzle.


\end{document}